\newcommand{\zlabel}[1]{\label{#1} }
\newcommand{\fc}{\frac} 
\newcommand{\lt}{\left} 
\newcommand{\rt}{\right} 
\newcommand{\mr}{\mathbf{r}}

\newcommand{\mx}{\mathbf{x}}

\newcommand{\al}{\alpha}

\newcommand{\pa}[2]{\frac{\partial #1}{\partial #2}}

\newcommand{\bu}{\mathbf{u}}

\newcommand{\lsc}{\;\text{\Large ;}}

\documentclass[%
 aip,
 amsmath,amssymb,
preprint,%
]{revtex4-1}

\usepackage{graphicx}
\usepackage{dcolumn}
\usepackage{epic}%
\usepackage{eepic}%
\usepackage{amsthm}
\usepackage{color}
\usepackage{bm}
\usepackage{comment}
\usepackage{esvect}
\usepackage{textcomp}

\begin{document}

\title{Developments of Bohmian Mechanics}

\author{James P. Finley}
\email{james.finley@enmu.edu}
\affiliation{
Department of Physical Sciences,
Eastern New~Mexico University,
Station \#33, Portales, NM 88130}
\date{\today}

\begin{abstract}
Bohmian mechanics is a deterministic theory of quantum mechanics that is based on a set of $n$
velocity functions for $n$ particles, where these functions depend on the wavefunction from the
$n$-body time-dependent Schr\"odinger equation.  It is well know that Bohmian mechanics is not
applicable to stationary states, since the velocity field for stationary states is the zero
function.
Recently, an alternative to Bohmian mechanics has been formulated, based on a conservation of
energy equation, where the velocity fields are not the zero function, but this formalism is
only applicable to stationary states with real valued wavefunctions.
In this paper, Bohmian mechanics is merged with the alternative to Bohmian mechanics. This is
accomplished by introducing an interpretation of the Bohm quantum potential. The final
formalism gives dynamic particles for all states, including stationary states. The final main
working equation contains two kinetic energy terms and a term that contains a factor that can
be interpreted as a pressure. The derivation is a simple $n$-body generalization of the recent
generalization, or refinement, of the Madelung equations.
\end{abstract}

\maketitle

\section{Introduction}

Bohmian mechanics\cite{Bohm:52a,Bohm:52b,B1,B2,B3,B4,B5,B6,B7,B8,B9,Jung,Renziehausenb} is a
deterministic theory of quantum mechanics that is based on a set of $n$ velocity functions for
$n$ particles, where these functions depend on the wavefunction from the $n$-body
time-dependent Schr\"odinger equation. The two equations of Bohmian mechanics are equivalent to
the time-dependent Schr\"odinger equation. An important function in Bohmian mechanics is the
Bohm quantum potential.\cite{Bohm:52a,Bohm:52b,B6}
It is well know that Bohmian mechanics is not applicable to stationary states, since the
velocity field for stationary states is the zero function.  
Therefore, as pointed out by Jung,\cite{Jung} a resting hydrogen $n$s electron would have a
strong non-zero electric dipole moment, and such a dipole moment would have been measured by
experiment, if it existed.
This same problem also appears in quantum hydrodynamics.\cite{B6}
Recently, an alternative to Bohmian mechanics has been formulated,\cite{Finley-ArxivDM} based on
a conservation of energy equation, where the velocity fields are not the zero function.
However, this method is only applicable to stationary states with real valued wavefunctions.
In this paper, Bohmian mechanics is merged with the alternative to Bohmian mechanics. This is
accomplished by introducing an interpretation of the Bohm quantum potential. The final
formalism gives a dynamic particles trajectories for stationary states. The final main working equation
contains two kinetic energy terms and a term that can be interpreted as a pressure. The
derivation is a simple $n$-body generalization of the recent generalization, or refinement, of
the Madelung equations.\cite{Finley-Made}

\section{Quantum Energy equation for Stationary States}

In this section we review the conservation of energy equation for quantum mechanical stationary
states with real valued wavefunctions,\cite{Finley-ArxivDM} where these equations are an
$n$-body generalization of fluid dynamic equations applicable to one-body
systems.\cite{Finley-Arxiv}

We also give a justification for the interpretation of the kinetic energy. The $n$-body
time-independent Schr\"odinger equation with a normalized, real-valued eigenfunction $R$ can be
written
\begin{equation} \zlabel{p2602}
  -\fc{\hbar^2}{2m}\sum_{i=1}^n \lt[R\nabla_i^2R\rt] + \sum_{i=1}^n V_i\Upsilon + \fc12 \sum_{i\ne j}^n R_{ij}^{-1}\Upsilon = \bar{E}\Upsilon,
\end{equation}
where 
\[
\lt[\psi\nabla_i^2\psi\rt](\mx) = \psi(\mx)\nabla_{\mr_i}^2\psi(\mx),  \quad \mx = \mathbf{x}_1,\cdots \mathbf{x}_n,
\]
and where the probability distribution is $\Upsilon = R^2$; also, the electron coordinate
$\mathbf{x}_i$ is defined by $\mathbf{x}_i = \mathbf{r}_i,\omega_i$, where
$\mathbf{r}_i\in\mathbb{R}^3$ and $\omega_i\in\{-1,1\}$ are the spatial and spin coordinates,
respectively. Furthermore, the $V_i$ and $R_{ij}^{-1}$ multiplicative operators are defined by
the following:
\[
[V_i\Upsilon](\mx) =  V(\mr_i)\Upsilon(\mx), \quad [R_{ij}^{-1}\Upsilon](\mx) = |\mr_i - \mr_j|^{-1}\Upsilon(\mx) \lsc  
\]
where the one-body external potential $V$ is a specified real-valued function with domain
$\mathbb{R}^3$ such that $\{\mr\in\mathbb{R}^3| R(\mx) = 0\}$ has measure zero. This last
requirements for $V$ implies that the division of a equation by $R$ or $\Upsilon$ gives an
equation that is defined almost everywhere.

Elsewhere\cite{Finley-ArxivDM} it is demonstrated that the Schr\"odinger equation (\ref{p2602}), with the
restriction $\Upsilon(\mx) \ne 0$, is equivalent to
\begin{equation} \zlabel{p4791b}
\sum_i \fc12 m u_i^2 + \Upsilon^{-1}\sum_i P_i +   \sum_{i=1}^n V_i + \fc12 \sum_{i\ne j}^n R_{ij}^{-1} = \bar{E}
\end{equation}
where 
\begin{gather} \zlabel{p4720}
  \bu_{i\pm}   = \pm\fc{\hbar}{2m}\fc{\nabla_i\Upsilon}{\Upsilon},
  \\ \zlabel{p4722} 
  P_i = -\fc{\hbar^2}{4m}\nabla_i^2\Upsilon
\end{gather}
and $u_i^2 = |\bu_{i\pm}|^2$. For the configuration $\mx = \mathbf{x}_1,\cdots \mathbf{x}_n$,
where one electron is at $\mx_1$, another one is located at $\mx_2$, and so on, the function
\[
\fc12 m [u_i(\mx_1,\mx_2,\cdots \mx_i,\cdots \mx_n)]^2
\]
is interpreted as the kinetic energy of the $i$th particle, i.e., the particle located at
$\mx_i$, where this particle has velocity $\bu_{i+}$ or~$\bu_{i-}$.
Equation~(\ref{p4791b}) can be interpreted as a classical energy equation with a Hamiltonian function
that depends on the probability distribution $\Upsilon$ and the potential energy functions
$V_i$ and $R_{ij}$.

For later use, we compare (\ref{p2602}) and (\ref{p4791b}), giving
\begin{equation} \zlabel{p2922}
-\fc{\hbar^2}{2m}\sum_{i=1}^n \lt[R\nabla_i^2R\rt] = \sum_i \fc12 \Upsilon m u_i^2  + \sum_i P_i 
\end{equation}

For each Cartesian coordinate $\al_i\in\{x_i,y_i,z_i\}$, $i = 1,\cdots n$, we require that the
wavefunction satisfy
\[
\lim_{\al_i \to \pm\infty} R(\mx) = \lim_{\al_i \to \pm\infty} \pa{R}{\al_i} = 0
\]
Hence
\[
\int_{-\infty}^{\infty} \pa{^2\Upsilon}{\al_i^2}\, d\al_i = \lt.\pa{\Upsilon}{\al_i}\rt|_{-\infty}^{\infty} =
2\lt.R\pa{R}{\al_i}\rt|_{-\infty}^{\infty} = 0 
\]
and therefore
\[
\int_{\mathbb{R}^3} \nabla_i^2\Upsilon\, d\mr_i = 0
\]
This result combined with (\ref{p4722}) gives
\[
\int P_i(\mr_i,\omega_i)\, d\mr_i = 0
\]
and it follows that the $P_i$ terms do not contribute to the expectation value of the kinetic
energy, denoted $\langle T\rangle$. Using the above equality for $P_i$, and integrating
(\ref{p2922}) over the $3n$ spatial coordinates and summing over the $n$ spin
coordinates, we have
\begin{equation} \zlabel{p4321}
\langle T\rangle = 
-\fc{\hbar^2}{2m}\sum_{i=1}^n \langle R|\nabla_i^2|R\rangle = \sum_i \fc12 m \langle R |u_i^2|R \rangle
\end{equation}
where Dirac notation and the definition $\Upsilon = R^2$ are used.  This result supports the
interpretations given above for the kinetic energy and velocity.

\section{Developed Bohmian Mechanics} 

The time-dependent Schr\"odinger equation is\cite{Levine,Bransden}
\begin{equation} \zlabel{p2588}
i\hbar\partial \Psi = -\fc{\hbar^2}{2m}\sum_i^n\nabla_i^2\Psi + U\Psi = \hat{H}\Psi \\ 
\end{equation}
where
\begin{equation} \zlabel{p0025}
U = \sum_{i\ne j}^n |\mr_i - \mr_j|^{-1} +  \sum_{i=1}^n V_i,
\end{equation}
$\Psi = \Psi(\mathbf{x},t)$ is the $n$-body time-dependent wavefunction, and we use the
same notation as in the previous section, e.g., $\mx_i = \mr_i,\omega_i$.  Let the spin
coordinates $\omega_i,\cdots \omega_n$ be specified parameters. Hence, $\Psi = \Psi(\mr,t)$,
where $\mr = \mr_1,\cdots \mr_n$, i.e., we can consider $\Psi$ to be a function of the spatial
coordinates and time only. For Bohmian mechanics,
\cite{Bohm:52a,Bohm:52b,B1,B2,B3,B4,B5,B6,B7,B8,B9,Jung,Renziehausenb} the wavefunction ansatz
\begin{equation} \zlabel{p4225}
\Psi = Re^{iSt/\hbar},
\end{equation}
where $R$ and $S$ are time-dependent real-valued functions, is substituted into the
time-dependent Schr\"odinger equation (\ref{p2588}), which, after significant manipulations,
yields the following two equations
\begin{gather} \zlabel{p4288}
  \partial \Upsilon + \sum_{i=1}^n\nabla\cdot(\Upsilon\bu_i)  = \mathbf{0}
  \\ \zlabel{p4299}  
  -R \partial S = \sum_{i=1}^n\lt(\fc{1}{2m}R \nabla_i S\cdot\nabla_i S - \fc{\hbar^2}{2m}\nabla_i^2 R  \rt) + UR,
\end{gather}
where the probability distribution is $\Upsilon = \Psi\Psi^* = R^2$ and $\partial S(\mx,t)=
\partial S(\mx,t)/\partial t$. Equation~(\ref{p4288}) is called the continuity equation. In the
special case of a one-body system, with $\Upsilon =\rho$, this equation has the same form as
the continuity equation from fluid dynamics,\cite{Munson,Shapiro} where the mass density is
$m\rho$. Bohmian mechanics, also defines the following two functions:
\begin{gather} \zlabel{p3312}
  \mathbf{v}_i = \fc{\nabla_iS}{m}
  \\ \zlabel{p3322}
  Q = -\fc{1}{R}\fc{\hbar^2}{2m}\nabla_i^2 R = \Upsilon^{-1}\lt(-R\fc{\hbar^2}{2m}\nabla_i^2 R\rt)
\end{gather}
where $ \mathbf{v}_i(\mx_1,\mx_2,\cdots \mx_i,\cdots \mx_n) $ is interpreted as the velocity of
the $i$th particle, i.e., the velocity of the particle located at $\mx_i$ for the configuration
$\mx = \mathbf{x}_1,\cdots \mathbf{x}_n$.  Also, $Q$ is known as the Bohm quantum
potential.\cite{Bohm:52a,Bohm:52b,B6} Substituting these two definitions into (\ref{p4299}),
and dividing by $R$, we get
\begin{equation} \zlabel{p5200}
 -\partial S = \sum_i\fc{1}{2}m v_i^2  + Q  + U
\end{equation}
where $v_i^2 = |\mathbf{v}_i|^2$.

Note that Eq.~(\ref{p2922}) is an equality holding for two times differentiable real-valued
functions, where $u_i^2 = |\bu_{i\pm}|^2$ and $P_i$ are given be Eq.~(\ref{p4720}) and
(\ref{p4722}), respectively.  Next we extend the interpretation of (\ref{p2922}) to the real
part of time-dependent wavefunctions $\Phi$, given by ansatz~(\ref{p4225}). Making this
interpretation and substituting Eq.~(\ref{p2922}) into (\ref{p3322}), we discover
\begin{equation} \zlabel{p5202}
Q = \sum_i \fc12 m u_i^2 + \Upsilon^{-1}\sum_i P_i
\end{equation}
The first term is a kinetic energy term. It is not necessary to interpret the second
term. However, one interpretation is the following:
$P_i(\mx_1,\mx_2,\cdots \mx_i,\cdots \mx_n)$
is the pressure experience by the $i$th particle, i.e., the particle located at $\mx_i$, for
the configuration $\mx = \mathbf{x}_1,\cdots \mathbf{x}_n$. Substituting (\ref{p5202})
into (\ref{p5200}) gives the desired result
\begin{equation} \zlabel{p5522}
-\partial S = \sum_i\fc{1}{2}m v_i^2  + \sum_i \fc12 m u_i^2 + \Upsilon^{-1}\sum_i P_i + U
\end{equation}
This equation is a further development of (\ref{p5200}), containing two kinetic energy terms, a
``compression'' energy term $\Upsilon^{-1}\sum_i P_i$, and the external potential $U$, given by
(\ref{p0025}).  The right-hand-side of this equation can be interpreted as the time dependent
total energy, i.e., a Hamiltonian function. For the left-hand side, from (\ref{p3312}), $S$ can
be interpreted as a momentum potential that is a sum of momentums from  all $n$ particles, but only including
the $\mathbf{v}_i$ portion of the total velocity $\bu_{i\pm} + \mathbf{v}_i$.

If $\Psi$ is a stationary state then $\Psi(\mx,t) = R(\mx)e^{-iEt/\hbar}$, so $S(t) = -\bar{E}t$, giving
\[
\sum_i\fc{1}{2}m v_i^2  + \sum_i \fc12 m u_i^2 + \Upsilon^{-1}\sum_i P_i + U = \bar{E}
\]
This equation is a generalization of Eq.~(\ref{p4791b}), holding for complex valued wavefunctions.

\section{Summary}
The Bohmian equation (\ref{p5200}), or (\ref{p4299}), with velocities given by (\ref{p3312})
and probability distribution $\Upsilon = \Psi\Psi^* = R^2$, is developed by an
interpretation the quantum potential, given by (\ref{p5202}), with velocities given by
(\ref{p4720}) and $P_i$ defined by (\ref{p4722}), giving the result (\ref{p5522}), with $U$
defined by (\ref{p0025}). The final velocities have been reinterpreted to be $\bu_{i\pm} +
\mathbf{v}_i$ instead of just $\mathbf{v}_i$. A justification for the additional velocity part
$\bu_{i\pm}$ is given by examining the expectation value of the kinetic energy (\ref{p4321}) in
the special case of a real-valued wavefunction $R$ of a stationary state.
The Lagrangian-function formulation by Salesi \cite{Salesi} and the generalized fluid-dynamics
formalism by Broer,\cite{Broer} come to the same conclusion, interpreting $\bu_{i\pm}$ as s
velocity.


\appendix

\bibliography{jfinley}

\end{document}